\documentclass[twocolumn,aps]{revtex4}
\usepackage{epsfig}
\newcommand{\beq}{\begin{eqnarray}}
\newcommand{\eeq}{\end{eqnarray}}
\begin{document}
\title{Hybrid Charmonium and the $\rho-\pi$ Puzzle}
\author{Leonard S. Kisslinger, Diana Parno and Seamus Riordan\\
Department of Physics, Carnegie Mellon University, Pittsburgh, PA 15213}

\begin{abstract} Using the method of QCD sum rules, we estimate the energy
of the lowest hybrid charmonium state, and find it to be at the energy of
the $\Psi'(2S)$ state, about 600 MeV above the $J/\Psi(1S)$ state. Since our
solution is not consistent with a pure hybrid at this energy, we conclude
that the $\Psi'(2S)$ state is probably an admixed $c \bar{c}$ and hybrid 
$c \bar{c}g$ state. From this conjecture we find a possible explanation of 
the famous $\rho-\pi$ puzzle.

\end{abstract}
\maketitle
\noindent
PACS Indices:14.40.Gx,12.38.Aw,11.55.Hx,13.25.Gv
\vspace{1mm}


\section{Introduction}

   A hybrid meson is a state composed of a quark and antiquark color octet,
along with a valence gluon, giving a color zero particle. Hybrids are
of great interest in studying the nature of QCD. The nonperturbative method 
of QCD sum rules~\cite{svz} has long been used to predict the energies of 
light quark hybrid mesons~\cite{gvgw} and hybrid baryons (see 
Ref.~\cite{kl95}). In the present work we use this method to estimate the 
energy of the lowest charmonium hybrid. A major motivation for the present 
work is to understand the nature of the $\Psi'(2S)$ state, and to find a
possible explanation of the long-standing $\rho-\pi$ puzzle.

  The $\rho-\pi$ puzzle concerns the branching ratios for hadronic decays of
the $\Psi'(2S)$ state compared to the $J/\Psi(1S)$ state. By taking ratios
of hadronic decays to gamma decays for these heavy quark states the
wave functions at the origin cancel, and by using the lowest-order
diagrams one obtains the ratios of branching rates for two charmonium states
\beq
\label{1}
  R&=&\frac{B(\Psi'(2S)\rightarrow h)}{B(J/\Psi(1S)\rightarrow h)}
\;=\;\frac{B(\Psi'(2S)\rightarrow e^+e^-)}{B(J/\Psi(1S)\rightarrow
e^+e^-)}\nonumber \\
         &\simeq& 0.13 \; ,
\eeq
the so-called 13 \% rule.
For the $\Psi'(2S)$ state compared to the $J/\Psi(1S)$ state,
however, the hadronic (e.g.$\rho-\pi$) decay ratio is more than an order of
magnitude smaller than predicted~\cite{markII}. This is the $\rho-\pi$ puzzle.
There have been many, many
theoretical attempts to explain this puzzle: Chen and Braatan~\cite{cb98}
review earlier work by Hou and Soni, Brodsky, Lepage and Tuan, Karl and
Roberts, Chaichian and Tornqvist, Pinsky, Brodsky and Karliner, and Li,
Bugg and Zou. All seem to agree that this is an unsolved puzzle. More
recently there has been an attempt to locate the source of the
problem~\cite{ms01}, with the suggestion that there is a cancellation of
two processes in the $\Psi'(2S)$ decay. Our present work suggests that one 
can obtain such a cancellation by including valence gluonic structure.

   In Sec. II we show that the energy of the lowest hybrid
charmonium state with the quantum numbers $J^{PC} =1^{--}$ is that of the 
$\Psi'(2S)$, but our solution is not consistent with a pure hybrid. Since the
$ c\bar{c} (2S)$ state is also expected to have that energy (about 600 MeV
above that of the $J/\Psi(1S)$ state), we predict that the $\Psi'(2S)$ state
is an admixture of $ c\bar{c} (2S)$ and hybrid components. In Sec. III we
show that this can provide a solution to the $\rho-\pi$ puzzle. In Sec. IV
we give our conclusions and compare our results to lattice gauge calculations.

\section{Hybrid $1^{--}$ Charmonium Using QCD sum rules}
   We now will use the method of QCD sum rules to attempt to find the lowest
hybrid charmonium state, assuming that such a pure hybrid charmonium meson
with quantum numbers $J^{PC}=1^{--}$ exists. First, let us review the method
and the criteria for determining if one has obtained a satisfactory and
accurate solution.

\subsection{Method of QCD Sum Rules for a Hybrid Charmonium Meson}

   The starting point of the method of QCD sum rules is the correlator, which
for a hybrid meson is
\beq
\label{2}
       \Pi^{\mu \nu }(x) &=&  \langle T[J^\mu_H(x) J^\mu_H(0)]\rangle \; ,
\eeq
with the current $J^\mu_H(x)$ creating the hybrid state being studied. The
QCD sum rule is obtained by evaluating $\Pi^{\mu \nu }$ in two ways. First, 
after a Fourier transform to momentum space, a dispersion relation gives the 
left-hand side (lhs) of the sum rule:
\beq
\label{3}
 \Pi(q)_{\rm{lhs}}^{\mu \nu} &=&  \frac{\rm{Im}\Pi^{\mu \nu }(M_A)}
{\pi(M_A^2-q^2)}+\int_{s_o}^\infty ds \frac{\rm{Im}\Pi^{\mu \nu }(s)}
{\pi(s-q^2)}
\eeq
where $M_A$ is the mass of the state $A$ (assuming zero width)
and $s_o$ is the start of the continuum--a parameter to be determined.
The imaginary part of $\Pi(s)$, with the term for the state we are
seeking shown as a pole (corresponding to a $\delta(s-M_A^2)$ term in 
$\rm{Im}\Pi$) and the 
higher-lying states produced by $J^\mu_H$ shown as the continuum, is 
illustrated in the figure:
\begin{figure}[ht]
\begin{center}
\epsfig{file=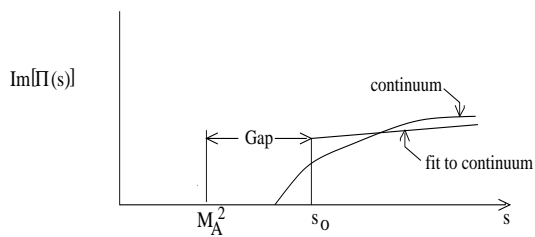,height=3cm,width=7cm}
\caption{QCD sum rule study of a state A with mass M$_A$ (no width)}
\label{Fig. 1}
\end{center}
\end{figure}

Next $ \Pi^{\mu \nu}(q)$ is evaluated by an operator product expansion
(O.P.E.), giving the right-hand side (rhs) of the sum rule
\beq
\label{4}
  \Pi(q)_{\rm{rhs}}^{\mu \nu} &=& \sum_k c_k(q) \langle 0|{\cal O}_k|0\rangle
 \; ,
\eeq
with increasing $k$ corresponding to increasing dimension of ${\cal O}_k$.

  After a Borel transform, ${\cal B}$, defined in  Appendix B,
in which the $q$ variable is replaced by
the Borel mass, $M_B$, the final QCD sum rule has the form
\beq
\label{5}
    && \frac{1}{\pi} e^{-M_{HHM}^2/M_B^2}
+ {\cal B} \int_{s_o}^\infty ds\frac{\rm{Im}[\Pi(s)]}{\pi(s-q^2)} \nonumber \\
     &=& {\cal B} \sum_k c_k(q) \langle 0|{\cal O}_k|0 \rangle \; ,
\eeq
as we shall show below.
Note that the Borel transform produces an exponential decrease with 
increasing values of $s$, as shown by the pole term in Eq(\ref{5}). This 
reduces the contribution of the continuum, where $s \geq M_B^2$.

This sum rule is used to estimate the heavy hybrid mass, $M_{HHM}$.
One of the main sources of error is the treatment of the continuum.
In addition to the parameter $s_o$, one must parameterize the effective
shape of the continuum. The criteria for a satisfactory solution are:
1) The contribution of the continuum should not be as large as the pole term
in the lhs. 2) With an exact sum rule, the value of $M_{HHM}$ is
independent of the value of $M_B$. With the approximation of a fit to the 
continuum, there should be a minimum or maximum in the value of 
$M_{HHM}$ vs $M_B$, and the value of $M_{HHM}$ at this extremum should
be approximately the value of $M_B$. 3) There should be a gap between
the solution for $M_{HHM}^2$ and $s_o$, which reduces the contribution 
of the continuum to be smaller than the pole term due to the factor of 
$e^{-s/M_B^2}$ after the Borel transform, as we shall explain below. 
If the value of $s_o$ is much larger than the expected excited hybrid 
states, however, the solution is not physical.

\subsection{Previous Results for Hybrids Using QCD Sum Rules and
Lattice QCD Methods}

  Since mesons with certain quantum numbers, such as $1^{-+}$, cannot
have a standard $q \bar{q}$ meson composition, there is a strong 
motivation for both experimental and theoretical searches for hybrid
mesons, which can have such states. These states are called exotic or
hermaphrodite mesons. Shortly after the introduction of QCD sum rules,
they were used to attempt to find the masses and widths of exotics.
The most accurate calculations~\cite{gvgw} predicted $1^{-+}$ hybrid
light-quark mesons in the 1.3-1.7 GeV region, where such a state has 
been found~\cite{page}. The solutions satisfy the criteria for a good 
solution, as explained above. For example,
for a solution that has a hybrid mass of 1.3 GeV, the value of $s_o$ was
1.7 GeV, which is a reasonable separation of the lowest from the higher
hybrid states. On the other hand, for these light meson hybrids many
terms in the O.P.E. are needed, which significantly increases the uncertainty.
This explains why previous calculations~\cite{gvgw} have found a rather wide
range of values for the light-quark exotic hybrid.

For heavy-quark hybrids the higher-order terms in the O.P.E. are quite
small, so the QCD sum rule method is more accurate. Since we are not 
trying to predict exotic hybrids, however, there is the serious 
complication of nonhybrid meson-hybrid meson mixing, which will be 
discussed in detail below. This is an important aspect of our present work.

   There have been many lattice QCD calculations of glueballs and 
light-quark hybrids~\cite{chen06}. The most recent calculations of light-quark
hybrids find the lightest exotics to be about 2 GeV~\cite{chen08}, quite a bit
higher in energy than the QCD sum rule calculations. This probably is due to
the fact that lattice QCD calculations for light quarks have  some 
inconsistencies at the present time, while they are much more accurate
for heavy quark systems~\cite{jkm}, as are QCD sum rules. Exotic charmonium
states have been calculated using lattice QCD, and the $1^{-+}$ was
found to be about 4.4 GeV~\cite{lm02}, with the expectation that the $1^{--}$
hybrid charmonium state is at a similar energy. For the calculation of 
nonexotic hybrid mesons, such as $1^{--}$ hybrids, there are 
other difficulties for both methods, as we shall discuss below.

\subsection{Heavy Hybrid Meson Correlator}

For a hybrid
meson with quantum numbers $1^{--}$ we use the standard current~\cite{gvgw}
\beq
\label{6}
         J^\mu_H &=&  \bar{\Psi}\Gamma_\nu G^{\mu\nu} \Psi \; ,
\eeq
with $\Gamma_\nu = C \gamma_\nu$, where C is the charge conjugation
operator, $\gamma_\nu$ is the usual Dirac matrix, and $\Psi$ is the 
heavy quark field. Carrying out a four-dimensional Fourier transform, 
the correlator in momentum space is
\beq
\label{7}
  \Pi^{\mu \nu }(p)& =& \int \frac{d^4 p_1}{(2 \pi)^4} Tr[S^{ab}
\Gamma_\alpha S^{ba} \Gamma_\beta](p-p_1) \nonumber  \\
  && Tr[G^{\mu \alpha} G^{\nu \beta})](p_1)  \; ,
\eeq
where $S^{ab}$ is the quark propagator, with colors $a$ and $b$. The
color properties of $G^{\mu \alpha}$, the gluon color field, with 
$\mu,\alpha$ Dirac indices, are given in Appendix A. Note that the
traces are both fermion and color traces. Details of $Tr[S^{ab}
\Gamma_\alpha S^{ba} \Gamma_\beta](p)$ and $Tr[G^{\mu \alpha} 
G^{\nu \beta})](p_1)$ are given in Appendix A.

   It is important to recognize that the correlator used in the QCD
sum rule method is similar to the correlator used in the lattice gauge
approach, with the same objective of finding the mass of a heavy hybrid
meson.

   As described in the preceding subsection, the correlator is evaluated
in the method of QCD sum rules via an O.P.E. and a dispersion relation.

\subsection{ O.P.E. of the Scalar Correlator in Momentum Space}

The QCD sum rule method uses an operator product expansion (O.P.E.) in
dimension (or inverse momentum), Eq(\ref{4}). For the hybrid meson the 
lowest-order diagram~\cite{fig} is shown in Fig. 2:

\begin{figure}[ht]
\begin{center}
\epsfig{file=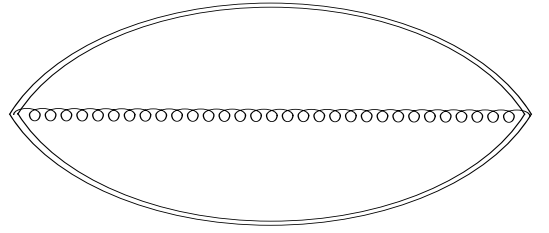,height=3cm,width=7cm}
\caption{ Lowest-order term in sum rule}
{\label{Fig.2}}
\end{center}
\end{figure}

Using the standard quark and gluon propagators (see Appendix A for some
details) we find for this lowest-dimensional process
\beq
\label{8}
   \Pi^{\mu \nu }_1(p) &=&  -6 \int \frac{d^4 p_1}{(2 \pi)^4}
\{g^{\mu \nu} \frac{[(p_1 \cdot (p-p_1)]^2}{p_1^2} \nonumber \\
  &&-\frac{p_1\cdot (p-p_1)}{p_1^2}[p^\mu p_1^\nu+p^\nu (p-p_1)^\mu
\eeq
\vspace{-3mm}
\beq
\label{9}
  && -2 p_1^\mu (p-p_1)^\nu]\}  \{\frac{2}{3} \frac{M_Q^2}{(p-p_1)^2}
 +[\frac{9}{4}(p-p_1)^2
\nonumber \\
  &&-\frac{41}{6} M_Q^2 -\frac{2}{3}M_Q^4 \frac{1}{(p-p_1)^2})] I_0(p-p_1)\}
\; , \nonumber
\eeq
with
\beq
\label{10}
      I_0(p) &=& \int_0^1\frac{d \alpha}{\alpha(1-\alpha) p^2 -M_Q^2} \; .
\eeq

   Extracting the scalar correlator $\Pi^S$, defined by
$\Pi^{\mu \nu}(p)=(p_\mu p_\nu/p^2-g^{\mu \nu} )\Pi^V(p) + (p_\mu p_\nu/p^2)
\Pi^S(p)$, and carrying out the $p_1$ integrals, one finds for $\Pi_1^S(p)$, 
 the scalar term of $\Pi^{\mu \nu }_1(p)$,
\beq
\label{11}
 \Pi_1^S(p) &=&-\frac{3}{(4\pi)^2}\int_0^1\frac{d\alpha}{\alpha(1-\alpha)}
\{\int_0^1 \frac{d\beta p^2 \beta}{p^2(1-\beta)-\frac{M_Q^2}{\alpha-\alpha^2}}
\nonumber
\eeq
\vspace{-3mm}
\beq
\label{12}
 && [-\frac{4}{3}\frac{M_Q^6}{\alpha-\alpha^2}-\frac{155}{12} \frac{M_Q^6}
{(\alpha-\alpha^2)^2}  +\frac{319}{12}p^2 \frac{M_Q^4}{\alpha-\alpha^2}
\nonumber \\
 &&-\frac{4}{3} M_Q^4 p^2-\frac{41}{3} M_Q^2 p^4]+  \\
  && \frac{p^4}{2}\int_0^1 d \beta \frac{-1}{p^2(1-\beta)-\frac{M_Q^2}
{\alpha-\alpha^2}}
[\frac{55}{3}M_Q^2(p^2 \nonumber \\
 &&-\frac{M_Q^2}{\alpha-\alpha^2})(-3\beta+ \beta^2- \beta^3/3)
  -\frac{8}{3} M_Q^4 (4 \beta  \nonumber \\
  && - 3 \beta^2+\beta^3/3)] \} \nonumber \\
  &&   + {\rm terms\;with\;three\;and\;four\;integrals} \nonumber \; .
\eeq
We find that the terms with three and four integrals are very small, and
we do not include them in our calculation.
\vspace{3mm}

The second term in the O.P.E for the heavy-quark hybrid
correlator includes the gluon condensate, illustrated in Fig. 3:

\begin{figure}[ht]
\begin{center}
\epsfig{file=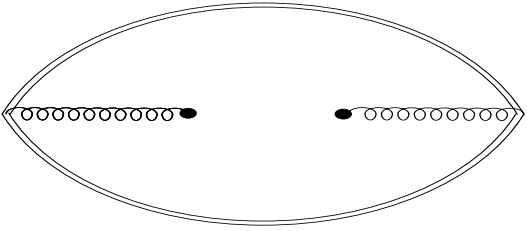,height=3cm,width=7cm}
\caption{Gluon condensate term in sum rule}
{\label{Fig.3}}
\end{center}
\end{figure}
For this process, the 
correlator $\Pi^{\mu \nu }_2(p)$ has the same form
as Eq.~(\ref{4}), except that the gluon trace used to obtain 
$\Pi^{\mu \nu }_1(p)$ is replaced with the trace over the gluon 
condensate~\cite{svz,gvgw}, which gives  
\beq
\label{13}
   Tr[G^{\mu \alpha} G^{\nu \beta}](p_1) &=& (2\pi)^4 \delta^4(p_1)
\frac{12}{96} <G^2>(g^{\mu \nu}g^{\alpha \beta} \nonumber \\
 &&-g^{\mu \beta}g^{\nu \alpha}) \; .
\eeq
The fermionic factor is the same as for Fig.~2. From this one finds
\beq
\label{14}
  \Pi^{\mu \nu }_2(p) &=& -\frac{3i}{2}\frac{<G^2>}{(4 \pi)^2}
\{ g^{\mu \nu}(-M_Q^2-\frac{p^2}{4}) \nonumber \\ 
  && (p^2-4 M_Q^2)I_0 +p^\mu p^\nu ( \frac{11}{2}p^2 -\frac{53}{3} M_Q^2
\nonumber \\
 &&+\frac{4}{3} \frac{M_Q^4}{p^2} ) I_0  +\frac{4}{3}\frac{M_Q^2}{p^2} \}
 \; ,
\eeq
giving for the scalar part of Fig.~3
\beq
\label{15}
  \Pi_2^S(p) &=& -\frac{3i}{2}\frac{<G^2>}{(4 \pi)^2}
(\frac{11}{4}p^4 \nonumber \\
  &&-\frac{59}{6} M_Q^2 p^2 +\frac{14}{3} M_Q^4 ) I_0 \; .
\eeq

\subsection{Borel Transform of the Correlator}

   To ensure convergence of the O.P.E. of the correlator, one performs a
Borel transform $\mathcal{B}$~\cite{svz}, defined in Appendix B. This is 
discussed in detail in the early papers on QCD sum rules\cite{svz,gvgw}. 
For our problem we have assumed that $\Pi^S(p) \simeq
\Pi_1^S(p) + \Pi_2^S(p)$, with higher-order terms being very small. We shall
see that even $\Pi_2$ is essentially
negligible within the accuracy of the method, and the convergence
after the Borel transform is established.

   Using the equations in Appendix B, from Eqs.~(\ref{B1},\ref{B2},\ref{B3}),
where the quantities $ B_1 $ through $ B_8 $ are given,
with $\mathcal{B}\Pi_1^S(p) \equiv \tilde{\Pi}_1^S(M_B)$,  we find

\beq
\label{16}
   \tilde{\Pi}_1^S(M_B) &=& -\frac{1}{4(4\pi)^2}
[-16 M_Q^6 B_1 -155 M_Q^6 B_2 \nonumber\\
&& +319 B_3 -16 M_Q^4 B_4 -164 B_5 \nonumber\\
&&-110(B_6- M_Q^2 B_7)-16 M_Q^4 B_8 ]
\eeq
\vspace{-4mm}
\beq
\label{17}
  &&= -\frac{1}{2(4\pi)^2}M_Q^4 \int_0^\infty d\delta
e^{-2 \frac{M_Q^2}{M_B^2}(1+\delta)}\nonumber \\
&&\{[-310 \frac{\delta}{1+\delta}+638 \delta -656 (1+\delta)] \nonumber \\
 &&K_3 (2\frac{M_Q^2}{M_B^2}(1+\delta) ) + [-1892 \frac{\delta}{1+\delta}+
\nonumber \\
&& 606 \delta -1968 (1+\delta) -32(4 \delta-3 \frac{\delta^2}{1+\delta}
\nonumber \\
&& + \frac{\delta^3}{3(1+\delta)^2})] K_2 (2\frac{M_Q^2}{M_B^2}(1+\delta) )
\nonumber \\
 && + [-4778 \frac{\delta}{1+\delta}+9442 \delta
-4920 (1+\delta) \nonumber \\
 &&-128(4 \delta-3 \frac{\delta^2}{1+\delta}
 + \frac{\delta^3}{3(1+\delta)^2})]\nonumber \\
&& K_1 (2\frac{M_Q^2}{M_B^2}(1+\delta) ) \\
 && + [-1356 \frac{\delta}{1+\delta}+6284 \delta -3280 (1+\delta) \nonumber \\
 &&-96(4 \delta-3 \frac{\delta^2}{1+\delta}+ \frac{\delta^3}{3(1+\delta)^2})]
 \nonumber \\
&&  K_0 (2\frac{M_Q^2}{M_B^2}(1+\delta) )+{\rm multiple\;integral\;terms}
 \; . \nonumber
\eeq
The multiple integral terms in Eq.(12) are small and are dropped, and $\delta$
is a variable of integration.

  In a similar way, taking the Borel transform of $\Pi_2^S(p)$, with
$ \mathcal{B} \Pi_2^S(p) \equiv \tilde{\Pi}_2^S(M_B)$, we find
\beq
\label{18}
   \tilde{\Pi}_2^S(M_B) &=&  -i\frac{3}{2(4\pi)^2}M_Q^4e^{-2 \frac{M_Q^2}
{M_B^2}}
\eeq
\vspace{-5mm}
\beq
 && [11 K_2 (2\frac{M_Q^2}{M_B^2} )+\frac{14}{3}K_1 (2\frac{M_Q^2}{M_B^2} )
 +18 K_0 (2\frac{M_Q^2}{M_B^2} )]  \nonumber\; .
\eeq

\subsection{QCD Sum Rule for Hybrid Charm Meson}

   The method of QCD sum rules uses a dispersion relation for the
correlator, which it equates to the correlator's operator product.
Following the usual convention we call the dispersion relation the left-hand 
side and the operator product expansion the right-hand side:
$\Pi_{\rm{lhs}}$ = dispersion relation, $\Pi_{\rm{rhs}}$ = operator product 
expansion. Neglecting the width of the hybrid meson, the dispersion relation 
has the form of a pole and a continuum: $\int_0^\infty ds\frac{\rm{Im}\Pi(s)}
{s-p^2}$.
The dispersion relation is evaluated in Euclidean space, $p^2\rightarrow -Q^2$,
and the continuum is assumed to start at $s=s_o$. After the Borel transform,
the form we use for the lhs is
\beq
\label{19}
    \tilde{\Pi}_{\rm{lhs}}(M_B^2)&=& F e^{-M_{HH}^2/M_B^2} +(L_1 M_B^2 
+ L_2 M_B^4)
\nonumber \\
  && \times  e^{-s_o/M_B^2} \; ,
\eeq
with $F$ the numerator of the pole, and $L_1$ and $L_2$ constants used to fit
the form of the continuum. We shall use a standard method to eliminate $F$,
and fit the sum rule requirement that the solution should not be sensitive
to $M_B$. In theory, the exact solution should be independent of $M_B$.

  For convenience in carrying out the sum rule, we have fit the rhs of the
correlator to a polynomial in the Borel mass, $ \tilde{\Pi}_1(M_B) = a_1 M_B^2
+a_2 M_B^4+a_3 M_B^6+a_4 M_B^8$ and $ \tilde{\Pi}_2(M_B) = b_1 M_B^2+b_2 M_B^4
+b_3 M_B^6+b_4 M_B^8$
with $a_1=122.29,\;a_2=-143.88,\;a_3=45.79,\;a_4=-0.346,\;b_1=616.0,\;b_2=
-279.1,\;b_3=-226.56,\;$and$\;b_4=144.515$, with units GeV$^6$.

   We find that, for all values of $M_B$ relevant to the sum rule, 
$\tilde{\Pi}_2$ is just about 1 \% of $\tilde{\Pi}_1$, and 
drop it, as the method is only valid to a few percent.
The sum rule is obtained by
taking the ratio of $\tilde{\Pi}$ to $\partial_{1/M_B^2} \tilde{\Pi}$. To do 
this we use the relations

\beq
\label{20}
  \partial_{1/M_B^2} e^{-\frac{A}{M_B^2}} K_\nu (-\frac{A}{M_B^2} )
\bigg\vert_{\nu>0}
&=& -Ae^{-\frac{A}{M_B^2}} K_\nu (-\frac{A}{M_B^2} ) \nonumber
\eeq
\beq
 && +\frac{A}{2}e^{-\frac{A}{M_B^2}}[ K_{\nu-1} (-\frac{A}{M_B^2} )
- K_{\nu+1} (-\frac{A}{M_B^2} )] \nonumber
\eeq
\beq
\label{21}
 \partial_{1/M_B^2} e^{-\frac{A}{M_B^2}} K_0 (-\frac{A}{M_B^2} )
&=&  -Ae^{-\frac{A}{M_B^2}}[ K_0 (-\frac{A}{M_B^2} ) \nonumber \\
&& - K_1 (-\frac{A}{M_B^2} )] \; .
\eeq

 Taking the ratio of $\partial_{1/M_B^2} \tilde{\Pi}_{\rm{lhs}}(M_B)=
\partial_{1/M_B^2} \tilde{\Pi}_{\rm{rhs}}(M_B)$ to the equation 
$\tilde{\Pi}_{\rm{lhs}}(M_B)=\tilde{\Pi}_{\rm{rhs}}(M_B)$,
one obtains the sum rule for the mass of the heavy charmonium hybrid meson:

\beq
\label{22}
 M_{HH}^2 &=& \{ e^{-\frac{s_o}{M_B^2}}[s_o(L_1 M_B^2+L_2 M_B^4)
+L_1 M_B^4 \nonumber \\
&&  +2 L_2 M_B^6] + \partial_{1/M_B^2} \tilde{\Pi}_1^S \} \\
 && \times \{ e^{-\frac{s_o}{M_B^2}}[(L_1 M_B^2 +L_2 M_B^4)
-\tilde{\Pi}_1^S \}^{-1} \nonumber \; .
\eeq

The result of the QCD sum rule fit is shown in Fig.~4, with $s_o$=60.0 
GeV$^2$, $L_1$=-99.0 GeV$^{^4}$, and $L_2$=6.06 GeV$^{2}$:

\begin{figure}[ht]
\begin{center}
\epsfig{file=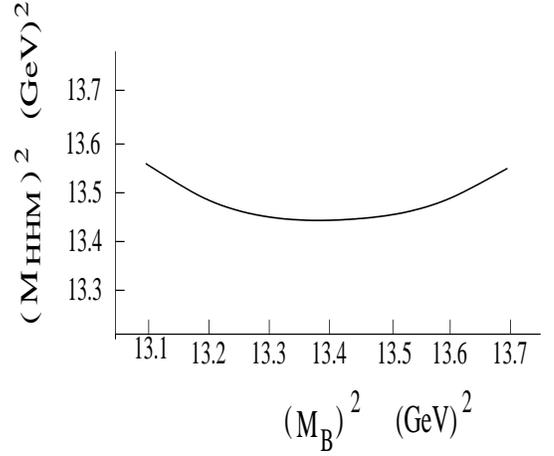,height=6cm,width=7cm}
\caption{QCD sum rule for heavy hybrid charm meson}
{\label{Fig.4}}
\end{center}
\end{figure}

   The parameters  $L_1$ and $L_2$ have magnitudes expected for the fit
to the continunmm.   Except for the large value of $s_o$,
the solution satisfies the criteria for a successful QCD sum rule: as Fig.~4 
shows, the solution for the mass is within a few per cent of the value of 
$M_B$ in the region of stability. The mass predicted for the hybrid is
\beq
\label{23}
        M_{HHM} &\simeq& 3.66 {\rm GeV} \; ,
\eeq
within a 10\% accuracy of the sum rule method,
while the experimental mass of the $\Psi'(2S)$ state~\cite{pdg} is
\beq
\label{24}
       M(\Psi'(2S)) &=& 3.68 {\rm GeV} \; .
\eeq
The large value of $s_o$, however, predicts that the excited hybrids are
at a very high energy, as found in lattice gauge calculations, and that
the $\Psi'(2S)$ cannot be a pure hybrid. Therefore, we expect that
the physical $\Psi'(2S)$ is an admixture with a charm meson and a hybrid
charm component. A second, orthogonal mixed state will be in the continuum.
As we now show, this can provide a solution to the $\rho-\pi$ puzzle.

\section{Hybrid-normal Charmonium and a Possible Solution to the $\rho-\pi$ 
Puzzle}

   In our treatment of hadronic decays of the $\Psi'(2S)$ state, we use the
Sigma/Glueball model, which was motivated by the BES analysis of glueball
decay~\cite{BES} and the study of scalar mesons and scalar glueballs using
QCD sum rules~\cite{kgv,kj}.  We briefly review this model below.

\subsection{The Sigma/Glueball Model}

   In energy regions where there are both scalar mesons and scalar
glueballs, it is expected that $0^{++}$ states will be an admixture of
mesons and glueballs. For this reason, when using QCD sum rules to find
such states one must use currents that are a linear combination of
glueball and meson currents. A scalar glueball current can have the form
\beq
\label{25}
      J^G (x) & = & \alpha_s G^2 \; ,
\eeq
while a scalar meson current has the form
\beq
\label{26}
     J^m (x) &  = & \frac{1}{2}(\bar{u} (x) u (x)- \bar{d} (x) d (x)) \; .
\eeq
We use for our $0^{++}$ current~\cite{kgv} (with $M_o$ needed for correct
dimensions) 
\beq
\label{27}
    J_{0^{++}} & = & \beta M_o J_m + (1-\mid\beta\mid) J_G \; .
\eeq

   The QCD sum rule calculation makes use of the correlator $ \Pi(x) =
 \langle T[J_{0^{++}} J_{0^{++}}] \rangle$. The cross term between $J^m$ 
and $J^G$ is evaluated by using the scalar glueball-meson coupling 
theorem~\cite{nov}:

\beq
\label{28}
     \int dx \langle T[J^G(x)J^m(0)] \rangle &\simeq& -\frac{32}{9}<\bar{q}q>
\; ,      
\eeq
with $ <\bar{q}q> \equiv {\rm the\;quark\;condensate}$ ,
which is illustrated in Fig. 5:
\begin{figure}[ht]
\begin{center}
\epsfig{file=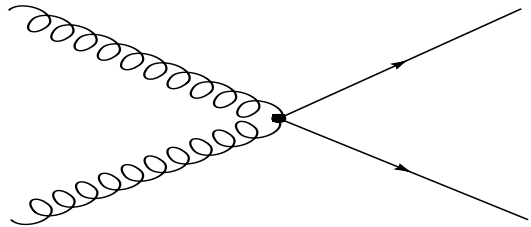,height=3.0cm,width=7cm}
\caption{Scalar glueball-meson coupling theorem}
{\label{Fig.5}}
\end{center}
\end{figure}

The results of the QCD sum rule calculations~\cite{kgv,kj} are that there are
three solutions:
\vspace{3mm}

  80\% scalar glueball at 1500 MeV $\rightarrow$ the f$_0$(1500)
\vspace{3mm}

80\% scalar meson at 1350 MeV $\rightarrow$ the f$_0$(1370)
\vspace{3mm}

Light Scalar Glueball 400-600 MeV $\rightarrow\;$ the  Sigma/Glueball
\vspace{3mm}

   The Sigma/Glueball model follows from the existence of the $\sigma$, a
scalar $\pi-\pi$ resonance with a broad width and the same mass as
the scalar glueball found in the sum rule calculations, and makes use of
the glueball-meson coupling shown in Fig. 5. It has been used for the
study of the Roper resonance decay into a nucleon and a $\sigma$~\cite{kl95,
kl99}, and the prediction of $\sigma$  production in proton-proton high
energy collisions~\cite{kms}, and other applications. The relevance for our
present work is that, just as it is possible to determine the mixing parameter
for a state consisting of a scalar meson and a scalar glueball, it should
be possible to determine the $c \bar{c}$ and hybrid $c \bar{c} g$ admixture
for charmonium systems, and for $\Upsilon$ systems.

\subsection{Hybrid Mixing Model and the $\rho-\pi$ Puzzle}

   The solution to the $\rho-\pi$ puzzle by the mixing of hybrid and
normal meson components of the $\Psi(2S)$ state can be understood from
Fig. 6 and Fig. 7, which illustrate the decay of a $c \bar{c}$ and a
$c \bar{c} g$ state into two hadrons. The $c \bar{c}$ decay
involves the matrix element $ <\pi \rho|O|\Psi'(c\bar{c},2S)>$:

\begin{figure}[ht]
\begin{center}
\epsfig{file=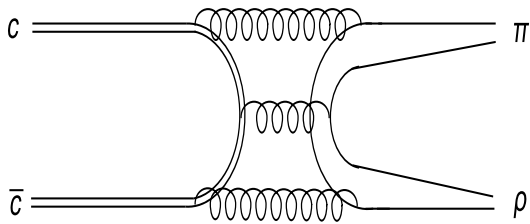,height=3.0cm,width=7cm}
\caption{Lowest-order PQCD diagram for a $c \bar{c}$ decay into two hadrons}
{\label{Fig.6}}
\end{center}
\end{figure}

The corresponding hybrid decay involves the matrix element
$<\pi \rho|O'|\Psi'(c\bar{c}g,2S)>$, with the diagram shown in Fig. 7.
\begin{figure}[ht]
\begin{center}
\epsfig{file=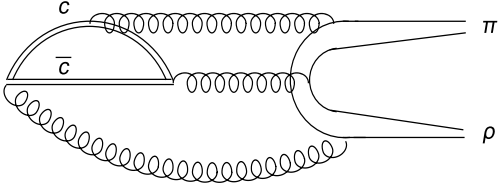,height=3.0cm,width=7cm}
\caption{Decay of a $c \bar{c} G$ state into two hadrons}
{\label{Fig.7}}
\end{center}
\end{figure}

Assuming that the 2s state is a $c\bar{c}$-$c\bar{c}g$ admixture
\beq
\label{29}
     |\Psi'(2S) \rangle &=& b|\Psi'(c\bar{c},2S) \rangle +\sqrt{1-b^2}
|\Psi'(c\bar{c}g,2S) \rangle
\nonumber \; ,
\eeq
and recognizing that the $O$ and $O'$ matrix elements are
approximately equal, we see that the solution to the $\rho-\pi$ puzzle requires
\beq
\label{30}
                 b + \sqrt{1-b^2} &\simeq& 0.1 \nonumber \; ,
\eeq
The solution of the $\rho-\pi$ puzzle would be given if
\beq
\label{31}
            b &\simeq -.7 \nonumber\; .
\eeq
In other words, if $b \simeq -.7 $, we have found a solution of the $\rho-\pi$ 
puzzle. The evaluation of $b$ is rather complicated, and the testing of this
conjecture will be carried out in future work.

\section{Conclusion}

   Using the method of QCD sum rules, we have shown that the $\Psi'(2S)$ state
cannot be a pure charmonium hybrid. We have found that the energy
of the lowest $J^{PC} =1^{--}$ hybrid charmonium state is approximately
the same as the $\Psi'(2S)$ state, about 600 MeV above the $J/\Psi(1S)$
state, but that the QCD sum rule solution is not consistent with a pure 
hybrid. The standard model prediction for $c \bar{c}(2S)$ is at approximately 
the same energy. Therefore we expect that the physical $\Psi'(2S)$ state is 
an admixture of a $c \bar{c}(2S)$ and a $c \bar{c}(8) g(8) (2S)$. Using 
this picture,we find a possible solution to the famous $\rho-\pi$ puzzle. 

   There have been many lattice calculations of exotic hybrid mesons. There
is experimental evidence for an exotic light-quark $1^{-+}$ meson (see 
Ref~\cite{page} for a discussion) at 1.4 to 1.6 GeV, which is consistent with 
QCD sum rule calculations~\cite{gvgw}, while lattice calculations find the 
lowest $1^{-+}$ hybrid at 1.9 to 2.1 GeV. The lowest energy $1^{-+}$ 
charmonium hybrid found in lattice calculations is at 4.4 GeV (see
 Ref~\cite{lm02}), about 800 MeV above our $1^{--}$ hybrid charmonium 
solution. This is consistent with our large value of $s_o$. For the $1^{--}$ 
state, we have shown that 
one must use a mixed $c\bar{c}$ and $c\bar{c}g$ current, which we 
shall use in future work. This use of a mixed current to define the 
correlator has not been done in lattice QCD caluclations, but with the QCD 
sum rule method it can be done in a rather straight-forward calculation.

   It is interesting that the energy difference between the $\Upsilon(2S)$
and the $\Upsilon(1S)$ is also approximately 600 MeV. If this is the energy of
a $\Upsilon(nS)$ hybrid, this could provide a solution
to the puzzling $\sigma$ decays of $\Upsilon(nS)$ states that have
recently been observed~\cite{vog06}. Investigation of this system is a 
topic of future research.

\section{Appendix}

\subsection{Heavy-Quark Hybrid Correlator in Momentum Space}

 The current to create a heavy-quark hybrid meson with $J^{PC} =1^{--}$ is
\beq
\label{A1}
         J^\mu_H &=&  \bar{\Psi}\Gamma_\nu G^{\mu\nu} \Psi \; ,
\eeq
where $\Psi$ is the heavy quark field, $\Gamma_\nu = C \gamma_\nu$,
$\gamma_\nu$ is the usual Dirac matrix, $C$ is the charge conjugation operator,
and the gluon color field is
\beq
\label{A2}
         G^{\mu\nu}&=& \sum_{a=1}^8 \frac{\lambda_a}{2} G_a^{\mu\nu}
\; ,
\eeq
with $\lambda_a$ the SU(3) generator ($Tr[\lambda_a \lambda_b]
= 2 \delta_{ab}$). From this one finds for the correlator of a heavy-quark
hybrid meson
\beq
\label{A3}
       \Pi^{\mu \nu }(p) &=& \int d^4x e^{i p\cdot x}
< T[J^\mu_H(x) J^\mu_H(0)]> \\
       &=& \int \frac{d^4 p_1}{(2 \pi)^4} Tr[S^{ab}
\Gamma_\alpha S^{ba} \Gamma_\beta](p-p_1) \nonumber  \\
  && Tr[G^{\mu \alpha} G^{\nu \beta})](p_1)  \; ,
\eeq
where $S^{ab}$ is the quark propagator, a standard Dirac propagator for
a fermion with mass $M_Q$, with colors $a$ and $b$. Using $C=i \gamma^2 \gamma^0$,
$Tr[\gamma_\alpha \gamma_\beta]=4g_{\alpha \beta}$, and $Tr[\gamma_\lambda
\gamma_\alpha \gamma_\delta \gamma_\beta]=4(g_{\lambda\alpha}g_{\delta\beta}
-g_{\lambda\delta}g_{\alpha\beta}+g_{\lambda\beta}g_{\delta\alpha})$:
\beq
\label{A4}
       Tr[S^{ab} \Gamma_\alpha S^{ba}\Gamma_\beta](p)&=& (-24 i)\int
\frac{d^D k}{(2\pi)^D}[-M_Q^2 g_{\alpha \beta}+ \nonumber
\eeq
\beq
 && (p_\lambda k_\delta-k_\lambda k_\delta)
(g_{\lambda\alpha}g_{\delta\beta}-g_{\lambda\delta}g_{\alpha\beta}
 +g_{\lambda\beta}g_{\delta\alpha})] \nonumber \\
 &&\frac{1}{(k^2-M_Q^2)[(p-k)^2-M_Q^2]} \; .
\eeq
To use dimensional regularization we define the quantity $D=4-\epsilon$,
and let $\epsilon \rightarrow 0$ to complete the integrals. One can then show
\beq
\label{A5}
     \int \frac{d^D k}{(2\pi)^D} \frac{1}{(k^2-M_Q^2)[(p-k)^2-M_Q^2]}
=\frac{I_0}{(8\pi)^2} \; ,
\eeq
with
\beq
\label{A6}
  I_0 &=& \int_0^1\frac{d\alpha}{(\alpha-\alpha^2)p^2-M_Q^2} \; .
\eeq

  The first term in the operator product expansion, shown in Fig. 2, has
a standard gluon propagator. For the gluon trace (see, e.g., Ref~\cite{gvgw})
one finds
\beq
\label{A&}
    Tr[G^{\mu \alpha} G^{\nu \beta}](p_1) &=& -4 \pi^2 i (g_{\alpha \beta}
\frac{p_1^\mu p_1^\nu}{p_1^2}+g_{\mu \nu}\frac{p_1^\alpha p_1^\beta}{p_1^2}
 \nonumber \\
&& -g_{\mu \beta}\frac{p_1^\alpha p_1^\nu}{p_1^2}
-g_{\alpha \nu}\frac{p_1^\mu p_1^\beta}{p_1^2} ) \; .
\eeq

The correlator for the process shown in Fig. 2 is found to be
\beq
\label{A8}
   \Pi^{\mu \nu }_1(p) &=&  -6 \int \frac{d^4 p_1}{(2 \pi)^4}
\{g^{\mu \nu} \frac{[(p_1 \cdot (p-p_1)]^2}{p_1^2} \nonumber \\
  &&-\frac{p_1\cdot (p-p_1)}{p_1^2}[p^\mu p_1^\nu+p^\nu (p-p_1)^\mu
\nonumber \\
  && -2 p_1^\mu (p-p_1)^\nu]\}  \{\frac{2}{3} \frac{M_Q^2}{(p-p_1)^2}
 +[\frac{9}{4}(p-p_1)^2
\nonumber \\
  &&-\frac{41}{6} M_Q^2 -\frac{2}{3}M_Q^4 \frac{1}{(p-p_1)^2})] I_0(p-p_1) \}
\; .
\eeq

The next term in the operator product expansion for this heavy quark system,
where quark condensates are negligible, is the gluon condensate term, shown
in Fig. 3. The trace over the quark propagators is the same as in
 Eq.~(\ref{A4}). The gluon field trace for this term is
\beq
\label{A9}
   <Tr[G^{\mu \alpha} G^{\nu \beta}](p_1)> &=& (2\pi)^4 \delta^4(p_1)
\frac{12}{96} <G^2> \nonumber \\
 &&(g^{\mu \nu}g^{\alpha \beta} -g^{\mu \beta}g^{\nu \alpha}) \; .
\eeq

From this one finds $\Pi_2^{\mu \nu}$, given in Eq.~(12).

\subsection{Borel Transforms}

   A key method that enables one to use the operator expansion to get
accurate sum rules is the use of the Borel transform~\cite{svz},
$\mathcal{B}$, defined by
\beq
\label{B0}
       \mathcal{B}= \lim_{q^2,n\rightarrow \infty}\frac{1}{(n-1)!}
(q^2)^n(-\frac{d}{d q^2})^n\bigg\vert_{q^2/n = M_B^2} \; .
\eeq

  Two key equations which we need are (with $K_\nu$ the modified
Bessel functions)
\beq
\label{B1}
        \mathcal{B}(\frac{1}{m^2-p^2})^k &=& \frac{e^{-m^2/M_B^2}}{(k-1)!
(M_B^2)^{k-1}} \nonumber \\
\label{B2}
        \int_0^\infty x^{\nu-1}e^{\frac{a}{x}-bx}&=& 2(\frac{a}{b})^{\nu/2}
K_\nu(2 \sqrt{ab}) \; .
\eeq

 Transforms used in the body of the paper are
\beq
\label{B3}
    \mathcal{B} I_0 &=& \int_0^1\frac{d\alpha}{(\alpha-\alpha^2)}
\mathcal{B}(p^2-M_Q^2/(\alpha-\alpha^2))^{-1} \nonumber \\
     &&= 2 e^{-2 M_Q^2/M_B^2}K_0 (2 M_Q^2/M_B^2 ) \nonumber \\
  B_1 &=& \mathcal{B} \int_0^1 \frac{d \alpha}{(\alpha-\alpha^2)^2} p^2
\int_0^1 \frac{d\beta \beta}{(1-\beta)[p^2-\frac{M_Q^2}{(\alpha-\alpha^2)
(1-\beta)}]} \nonumber \\
  &=& 2 M_Q^2\int_0^\infty d \delta \frac{\delta}{1+\delta}
e^{-2 \frac{M_Q^2}{M_B^2}(1+\delta)}[2 K_2 (2 \frac{M_Q^2}{M_B^2}(1+\delta))
\nonumber \\
&& +8K_1 (2\frac{M_Q^2}{M_B^2}(1+\delta))+6K_0(2 \frac{M_Q^2}{M_B^2}
(1+\delta))]
  \nonumber \\
   B_2 &=& \mathcal{B} \int_0^1 \frac{d \alpha}{(\alpha-\alpha^2)^3} p^2
\int_0^1 \frac{d\beta \beta}{(1-\beta)[p^2-\frac{M_Q^2}{(\alpha-\alpha^2)
(1-\beta)}]} \nonumber \\
  &=& 2 M_Q^2\int_0^\infty d \delta \frac{\delta}{1+\delta}
e^{-2 \frac{M_Q^2}{M_B^2}(1+\delta)}[ 2 K_3(2 \frac{M_Q^2}{M_B^2}(1+\delta))
\nonumber \\
&& +12K_2(2\frac{M_Q^2}{M_B^2}(1+\delta))+30K_1(2\frac{M_Q^2}{M_B^2}
(1+\delta)) \nonumber \\
  &&+20K_0(2 \frac{M_Q^2}{M_B^2}(1+\delta))] \nonumber \\
  B_3 &=& \mathcal{B} \int_0^1 \frac{d \alpha}{(\alpha-\alpha^2)^3} p^4
\int_0^1 \frac{d\beta \beta}{(1-\beta)[p^2-\frac{M_Q^2}{(\alpha-\alpha^2)
(1-\beta)}]} \nonumber \\
   &=& 2 M_Q^4\int_0^\infty d \delta \delta
e^{-2 \frac{M_Q^2}{M_B^2}(1+\delta)}[ 2 K_3(2 \frac{M_Q^2}{M_B^2}(1+\delta))
 \nonumber \\
 && +12K_2(2\frac{M_Q^2}{M_B^2}(1+\delta))+30K_1(2\frac{M_Q^2}{M_B^2}
(1+\delta))] \nonumber \\
  && +20K_0(2 \frac{M_Q^2}{M_B^2}(1+\delta))] \nonumber \\
  B_4 &=& \mathcal{B} \int_0^1 \frac{d \alpha}{(\alpha-\alpha^2)} p^4
\int_0^1 \frac{d\beta \beta}{(1-\beta)[p^2-\frac{M_Q^2}{(\alpha-\alpha^2)
(1-\beta)}]} \nonumber \\
  &=&  2 M_Q^4\int_0^\infty d \delta \delta
e^{-2 \frac{M_Q^2}{M_B^2}(1+\delta)}[ 2 K_2(2 \frac{M_Q^2}{M_B^2}(1+\delta))
 \nonumber \\
  B_5 &=& \mathcal{B} \int_0^1 \frac{d \alpha}{(\alpha-\alpha^2)} p^6
\int_0^1 \frac{d\beta \beta}{(1-\beta)[p^2-\frac{M_Q^2}{(\alpha-\alpha^2)
(1-\beta)}]} \nonumber \\
 &=&  2 M_Q^6\int_0^\infty d \delta \delta (1+\delta)
e^{-2 \frac{M_Q^2}{M_B^2}(1+\delta)}[ 2 K_3(2 \frac{M_Q^2}{M_B^2}(1+\delta))
\nonumber \\
&& +12K_2(2\frac{M_Q^2}{M_B^2}(1+\delta))+30K_1(2\frac{M_Q^2}{M_B^2}
(1+\delta))] \nonumber \\
  &&+20K_0(2 \frac{M_Q^2}{M_B^2}(1+\delta))] \nonumber \\ 
 B_6 &=& \mathcal{B} \int_0^1 \frac{d \alpha}{(\alpha-\alpha^2)} p^6
\int_0^1 \frac{d\beta (-3\beta+\beta^2-\beta^3/3)}{(1-\beta)[p^2-\frac{M_Q^2}
{(\alpha-\alpha^2)(1-\beta)}]} \nonumber \\
 &=&  2 M_Q^6\int_0^\infty d \delta (-3\delta+\frac{\delta^2}{1+\delta}-
 \frac{\delta^3}{3(1+\delta)^2)}) e^{-2 \frac{M_Q^2}{M_B^2}(1+\delta)} 
\nonumber \\
&&[ 2 K_3(2 \frac{M_Q^2}{M_B^2}(1+\delta)) +12K_2(2\frac{M_Q^2}{M_B^2}
(1+\delta)) \nonumber \\
&&+30K_1(2\frac{M_Q^2}{M_B^2}(1+\delta))+20K_0(2 \frac{M_Q^2}{M_B^2}
(1+\delta))] \nonumber \\
  B_7 &=& \mathcal{B} \int_0^1 \frac{d \alpha}{(\alpha-\alpha^2)^2} p^4
\int_0^1 \frac{d\beta (-3\beta+\beta^2-\beta^3/3)}{(1-\beta)[p^2-\frac{M_Q^2}
{(\alpha-\alpha^2)(1-\beta)}]} \nonumber \\
  &=& \frac{1}{M_Q^2} B_6 \nonumber 
\eeq
\newpage

\beq
 B_8 &=&  \mathcal{B} \int_0^1 \frac{d \alpha}{(\alpha-\alpha^2)} p^4
\int_0^1 \frac{d\beta (4\beta-3\beta^2+\beta^3/3)}
{(1-\beta)[p^2-\frac{M_Q^2}{(\alpha-\alpha^2)
(1-\beta)}]} \nonumber \\
 &=&2 M_Q^4\int_0^\infty d \delta (4\delta
-\frac{3 \delta^2}{1+\delta}+\frac{\delta^3}{3(1+\delta)^2}))\nonumber \\
&&e^{-2 \frac{M_Q^2}{M_B^2}(1+\delta)}[ 2 K_2(2 \frac{M_Q^2}{M_B^2}(1+\delta))
 \nonumber \\
&& +8K_1(2\frac{M_Q^2}{M_B^2}(1+\delta))
+6K_0(2 \frac{M_Q^2}{M_B^2}(1+\delta))] .
\eeq
\vspace{3mm}

\large{{\bf Acknowledgments}}
\normalsize
\vspace{1mm}

  This work was supported in part by the NSF/INT grant number 0529828.

  The authors thank Professors Pengnian Shen, Wei-xing Ma, and other
IHEP, Beijing colleagues for helpful discussions. We thank Professor Y. Chen
for discussions of lattice QCD in comparison to QCD sum rules for hybrid
states.


\begin{thebibliography}{99}
\bibitem{svz}M.A. Shifman, A.I. Vainstein and V.I. Zakharov, Nucl Phys.
{\bf B147} 385, 448 (1979)
\bibitem{gvgw}J. Govaerts, F. de Viron, D. Gusbin and J. Weyers, Nucl. Phys.
{\bf B248} 1 (1984); J.I. Latorre, S. Narison, P. Pascual and R. Tarrach,
Phys. Lett. {\bf B147}, 169 (1984). Contain references to earlier work
\bibitem{kl95} L.S. Kisslinger and Z. Li, Phys. Rev. {\bf D51} R5986 (1995)
\bibitem{markII} M.E.B Franklin et. al. (Mark II Collaboration), Phys. Rev.
Lett. {\bf 51} 963 (1983)
\bibitem{cb98}Y-Q. Chen and E. Braaten, Phys. Rev. Lett. {\bf 80} 5060 (1998)
\bibitem{ms01}M. Suzuki, Phys. Rev. {\bf D63} 054021 (2001)
\bibitem{page}P.Page, arXiv:hep-ph/9909201
\bibitem{chen06} Y. Chen et al, Phys. Rev. {\bf D 73} 014516 (2006), contains
references to earlier work
\bibitem{chen08} Y. Chen, private communiation
\bibitem{jkm}K.J. Juge, J.Kuti and C.J. Morningstar, Nucl.Phys. Proc. Suppl.
{\bf 83} 304 (2000)
\bibitem{lm02}X.Liao and T. Manke, arXiv:hep-lat/0210030
\bibitem{fig} D. Binos and L. Theu{\ss}l, Computer Phys. Communications
{\bf 161}, 76 (2004)
\bibitem{BES} L.Y. Dong (BES Collaboration) BES CONF97
\bibitem{kgv} L.S. Kisslinger, J Gardner and C. Vanderstraeten, Phys. Lett.
{\bf B 410}, 1 (1997)
\bibitem{kj} L.S. Kisslinger and M.B. Johnson, Phys. Lett. {\bf B 523}, 127
(2001)
\bibitem{nov} V.A. Novikov, M.A. Shifman, A.I. Vainstein, V.I. Zakharov, 
Nucl. Phys. {\bf B165} 67 (1980); Nucl. Phys. {\bf B191} 301 (1981)
\bibitem{kl99}L.S. Kisslinger and Z. Li, Phys. Lett. {\bf B445}, 271 (1999)
\bibitem{kms} L.S. Kisslinger, W-h. Ma and P. Shen, Phys. Rev. {\bf D71},
094021 (2005)
\bibitem{pdg} Review of Particle Physics, W-M. Yao et al, J. Phys. {\bf G 33},
1 (2006).
\bibitem{vog06}H. Vogel, FPCP 2006, Vancouver, BC, arXiv:hep-ex/0606011.
 \end{thebibliography}
\end{document}